\documentclass[12pt]{article}
\usepackage{amsthm,amsfonts,amssymb,latexsym}
\usepackage{color}
\begin{document}

\title{The present situation in quantum theory and its merging with general relativity}

\author{Andrei Khrennikov\\
International Center for Mathematical Modelling \\
in Physics and Cognitive Sciences\\
Linnaeus University,
V\"axj\"o, SE-351 95, Sweden\\
National Research University of  Information Technologies\\
Mechanics and Optics (ITMO), St. Petersburg 197101, Russia\\
Andrei.Khrennikov@lnu.se}


\maketitle

\abstract{We discuss the problems of quantum theory (QT) complicating  its merging with general relativity (GR). 
QT is treated as a general  theory of  micro-phenomena - a bunch of  models. Quantum mechanics (QM) and quantum field theory (QFT) are the most widely known  (but, 
e.g., Bohmian mechanics is also a part of QT). The basic problems  of QM and QFT are considered in interrelation.  
For QM, we stress its nonrelativistic character and the presence of spooky action at a distance. For QFT, we highlight the old problem of infinities. 
And this is the main point of the paper: it is meaningless to try to unify QFT so heavily suffering of infinities with GR.  
We also highlight difficulties of the QFT-treatment of entanglement.
We compare the QFT and QM based measurement theories by presenting both theoretical and experimental viewpoints. 
Then we discuss two basic mathematical constraints of  both QM and QFT, namely,  the use of real (and, hence, complex) numbers and the Hilbert state space. We briefly present non-Archimedean and non-Hilbertian 
approaches to QT and their consequences.  Finally,  we claim that, in spite of the Bell theorem, it is still possible to  treat quantum phenomena  
on the basis of  a classical-like causal theory.  We  present a random field model generating  the QM and QFT  formalisms.
This emergence viewpoint can serve as the basis for unification of novel QT (may be totally different from presently powerful  
QM  and QFT) and general relativity GR. (It may happen that the latter would also be revolutionary modified.)}

{\bf  keywords:} quantum theory, quantum mechanics, quantum field theory, quantum gravity, infinities, measurement theory, non-Hilbertian state space formalism, 
non-real numbers in physics, emergent theory

\section{The main ideas of this paper}

We start this paper with a brief presentation of its main ideas (the talk of the author at the conference, 
``From Foundations of Quantum Mechanics to Quantum Information and Quantum Metrology and Sensing'', 
Turin, May 11, 2017). The main body of the paper starts with section 2.

\subsection{QM versus QFT}

Analysis of  the fundamental problems of modern QT. Some of them  as the main barrier preventing unification of QT and GR.

Two basic QT-models, QM and QFT, have some degree of independence - in general ``they live their own lives''.  
It is commonly stated:  QFT is the fundamental theory and QM is its  non-relativistic approximation.

Is QM just a technical tool for  calculations in the non-relativistic limit? It seems that this is  not the case. 

QM is not just a computational device. The most exciting modern  foundational discussions, especially those related to the recent quantum information revolution,  are mainly 
going in the QM-framework. Why?

My opinion: quantum measurement theory is well established only in the QM-framework! 

 It is difficult  to handle the concrete measurement in the rigorous QFT-framework. 
E.g., the most intriguing problem of modern quantum foundations, the problem of  interrelation 
of local realism and QT, was treated (with a very few exceptions) only in  QM. 

{\bf  {\color{green} Entanglement: can it be adequately treated in QFT, by being frame-dependent?}}

At the same time QM cannot function without QFT: QFT  secures that QT does not contradict to special relativity. 

The only mathematically and conceptually sound QM is nonrelativistic QM. 

"Relativistic QM" suffers of a number of difficult problems, e.g., the problem of relativistic localization (position).  By  Hegerfeldt's theorem:  
{\it Einstein causality implies that there can be no spectral measure solution to the localization problem in relativistic QM.}

\subsection{Nonlocality mess}

QT as QM is nonlocal: action at a distance,  violation of Bell's inequality and "quantum nonlocality".

QT as QFT is local and relativistically invariant.

The notion of locality in QM and QFT have very different meanings!

Remark:  we are excited by having the QM-nonlocality, but afraid of nonlocal versions of QFT.

\subsection{The problem of infinities}

Main problem of  QFT:  {\it divergences.} Although it
is well known, nowadays it is hidden ``under the table'': development of advanced techniques of renormalization made 
the impression that QFT is a consistent theory.

{\bf  {\color{blue} There are no infinities in the physical world and  infinities are just the fruits of the human mind.}} 
 The appearance of infinity is just a sign that some theory is applied outside its real domain of application.

  ``Late Dirac'' was not satisfied by QFT and also because of the presence of infinities.
In the early 1980s, P. Dirac told E.  Witten that the most important challenge in physics was {\bf  {\color{red}"to get rid of infinity!"}} 

Dirac: the main problem of QFT is in formal treatment of the vacuum state  $\vert 0 \rangle.$ The real physical vacuum has a complex structure
and this structure cannot be reduced to operation with the symbol $\vert 0 \rangle.$ 
 Dirac's attempt to reorganize QFT by clarifying the role of vacuum, can be treated as a step towards quantum gravity.

\subsection{ On merging QT with GR}

We should not try to unify QT suffering of divergences (at the QFT level) with GR. First we have to resolve the biggest, but 
commonly ignored (nowadays) problem of QT, namely, creation of QFT without infinities and their handmade regularizations. 

{\bf  {\color{red} Where is Devil?}}

a) {\bf  {\color{blue} in Hilbert space!}} (Dirac, Khrennikov, and recently `t Hooft)

b) {\bf  {\color{blue} in the real continuum!}} (Manin, Volovich, Vladimirov, Witten, Framton, Dragovich, Khrennikov and recently `t Hooft) 

c) {\bf  {\color{blue} in belief in completeness of QM}} (Einstein, `t Hooft, Khrennikov)

`t Hooft, G.: The nature of quantum mechanics. Conference The Future of Physics, Santa Barbara, 2004.\\
  http://online.itp.ucsb.edu/online/kitp25/thooft/oh/01.html
 
A. Khrennikov, The Present Situation in Quantum Theory and its Merging with General Relativity, Found. Phys., 2017 (Open access) or arXiv.

\subsection{ $P$-adic space project}

The only "physical numbers" are rational. However, to have analysis,  $\bf {Q}$ has to be completed. 
One completion is the field of real numbers $\bf {R}.$ Are there other completions of $\bf {Q}?$
(with extension of arithmetics from $\bf {Q}).$ {\bf  {\color{blue} Ostrovsky theorem}} of number theory: only field of $p$-adic numbers
$\bf {Q}_p,$ based on prime numbers $p>1.$  Let us try to make theoretical physics not only with $\bf {R},$ but even 
with $\bf {Q}_p?$ Which p? All! Adelic approach.

QFT: B. Dragovich, A.   Khrennikov, S. V. Kozyrev and I. V. Volovich, On p-adic mathematical physics {\it P-Adic Numbers, Ultrametric
Analysis, and Applications}, {\bf  1}, N 1, 1-17 (2009).

The main feature of $\bf {Q}_p:$  it is totally disconnected,  each ball can be split into $p$ disjoint balls.  

Divergences: no ultraviolet divergences, but infrared are still present! So, the change of the basic number field from the real numbers to p-adic 
does not solve the problem of divergences. 

\subsection{Einstein's dream}

 QT from classical field theory: 
 
{\bf  {\color{red} 
A. Einstein and L. Infeld,   Evolution of Physics: The Growth of Ideas from Early Concepts to Relativity and Quanta. Simon  and Schuster, New-York (1961).}}

By distancing from statements such as the Bell no-go theorem we can consider the following plan for unification of QT with GR \cite{KH}:
\begin{itemize}
\item Unification of all physical interactions in a single  classical field model ${\cal M}.$ 
\item Emergence of QT from  ${\cal M}.$
\end{itemize}
Of course, ${\cal M}$ has to be local and to have nothing to do with nonlocal subquantum models of the Bohmian type

\subsection{ What is about "no-go theorems"?}

One may immediately say that such  a model ${\cal M}$ does not exist as the result of the experimental confirmation of a violation of the Bell inequality. 
However, this conclusion depends on treatments of possible ways of coupling a subquantum (ontic) model with QM (epistemic model).
In short, a violation of the Bell inequality closed only one special way for coupling of subquantum and quantum models.

Remind: The first no-go theorem was von Neumann's theorem.  It was based on different set of constraints. Bell started with critique of von Neumann's assumptions and relaxed
them. One can relax Bell's assumptions and still have physically "natural" theory. 

A. Khrennikov,  {\bf  {\color{green} After Bell.}} Fortschritte der Physik- Progress in Physics (2016). \\
http://onlinelibrary.wiley.com/doi/10.1002/prop.201600044/epdf 

One  can still  treat QM as emergent from a causal classical-like model ${\cal M}.$
For us, it is important  that  ${\cal M}$ can be a classical field model. 
\subsection{ Prequantum classical statistical field theory (PCSFT)}

 Khrennikov,  A.:  Beyond Quantum.  Pan Stanford Publ., Singapore (2014).
 The idea behind this model and emergence of QM is very simple: 

{\bf  {\color{red} Quantum density operators are just the symbolic representations of covariance operators classical random fields.}}

Such fields fluctuate at the time and space scales which are essentially finer than the scales of quantum experimental physics.
We are not able to monitor individual fluctuations; only correlations can be measured. 

There is a kind of nonlocal element in PCSFT. However, this is not nonlocality of mystical action at a distance, but the presence of correlations 
in the {\bf  {\color{blue} common background field}} feeling the space-time (similar to the zero-point field). 
Its correlations contribute into the ``quantum correlations.'' We repeat once again that 
correlations in such a background field are typical for, e.g., radio-engineering.

\subsection{Summary}

 \begin{enumerate}
\item Nonrelativistic QM versus relativistic QFT were just steps towards real QT which have not yet been constructed.
\item QFT suffers of infinities. Ignorance of infinities under the shadow of regularization procedures cannot be considered as acceptable.
\item Infinities and singularities  are foreign to real physical phenomena. 
A theory containing singularities cannot be considered as physically adequate, neither QFT nor Big Bang. 
\item Foundational output of violation of Bell's inequality - existence of action at a distance, has to be translated to the QFT-language.
\item The latter seems to be impossible, since there is no the QFT-based  notion of entanglement, it is frame dependent.
\item Unification  of QT with GR is a very risky adventure, because the real QT has not yet been created. 
Its initially created part in the form of QM suffers of the absence of relativistic invariance, it advanced part, QFT, suffers of infinities  and problems with 
relativistic notion of entanglement. 
\end{enumerate}

\section{Detailed introduction}

The last year  a few  experimental groups  performed  the loophole free tests of violation of the Bell inequality \cite{B1, B2}, see  \cite{E1}-\cite{E3}.  
This is really an exciting event in the quantum foundations! Leading experts in quantum foundations consider these tests as the final point 
in the exciting debate about the EPR-experiment, see, e.g., \cite{AF}, \cite{W} (cf., however, \cite{KH}, \cite{KP}). In 
this euphoric  situation the only commonly recognizable problem of  the quantum community is the problem 
of merging QT (in any its form) with GR.  The conventional viewpoint is that there is only one 
black cloud on the blue quantum sky: the evident impossibility to unify QT with GR.\footnote{This viewpoint was presented in numerous discussions of the author  
with  experts in QT. Of course, they know well about 
other problems of QT. But all such problems are treated as technicalities which can be solved in future, see also appendix.} 
Does this picture reflect correctly the present situation in QT? 

In this paper we shall analyze the fundamental problems of modern QT. We consider some of them  as the main barrier preventing unification of QT and GR.\footnote{There has been some 
valuable work done in this direction, see,  in particular, \cite{QFTG}.} We point out that, in fact,
two basic QT-models, QM and QFT, have some degree of independence - in general ``they live their own lives''.  
It is commonly stated, see, e.g., citations in appendix, that QFT is the fundamental theory and QM is its  non-relativistic approximation.
To be consistent with this viewpoint, we have to reduce the value of QM  and treat it  as 
a technical tool useful for concrete calculations in the non-relativistic limit. However, this is really not the case. 

The indirect sign of the impossibility of diminishing the role of  QM to the role of a non-relativistic computational tool  is that  the most exciting modern 
foundational discussions, especially those related to the recent quantum information revolution,  are mainly 
going in the QM-framework.\footnote{ Only recently  quantum information 
was coupled with QFT-foundations \cite{DA1}-\cite{PD1}.
In general foundational QFT-studies are not so widely present at the quantum arena (see, however, 
e.g., \cite{QFT1}- \cite{QFT4} ), see also appendix: QM versus QFT foundations.}

To complete this discussion on the role played by research on QM-foundations in general, 
 I remark that it is quite common for experimentalists to claim  that modern QM-foundational debates are of the minor value for real physics:  QFT (which is claimed to be 
the real physical theory) is free of all 
QM-ambiguities - the relicts of the first period of development of  QT.  
Surprisingly, this viewpoint is shared even by some experimentalists doing 
experiments related to quantum foundations. They like ``cream on the top of foundational cappuccino'' (as the Bell inequality), but at the same time they do not want to drink 
the rest: collapse of the wave function, the Born rule, the measurement problem and measurement theory (POVMs, theory of open quantum systems). 
Of course, this attitude  is surprising for 
a theoretician. However, an experimentalist may really think that all these modern studies about the Born rule (e.g., where did it come from?) or quantum instruments are useless. 
There are detectors and there are clicks in them. Then the task of an experimentalist is to count the numbers of clicks in different detectors (and to control the level of noise, loopholes and other 
``technicalities'').   We remark that this position matches well with the Copenhagen interpretation of QM and the views 
of N. Bohr:  QM is about prediction of outcomes of classical measurement devices.\footnote{In fact, Bohr's concept of measuring devices is more complex.  
What Bohr says is that the observable parts of measuring instruments can be
described by classical physics, while they also have quantum parts by means of
which they interact with quantum objects.}   This attitude of experimentalists 
to neglect the achievements of modern theory of quantum measurements  is interesting to analyze and we shall discuss it in more detail in section \ref{MT}.

Personally I think that one cannot proceed with analysis of experiments without a theory of quantum measurements. And the latter 
was consistently developed (at the mathematical level of rigorousness)  mainly in the QM-framework,
namely, as theory of POVMs-based quantum instruments \cite{Davies}-\cite{PB}, see also  \cite{Jaeger3} on discussion about  interpretational 
issues.\footnote{For the QFT-measurement theory, see, e.g., Schweber \cite{Schweber}, Schwinger \cite{Schwinger};  from the foundational viewpoint,
 the pioneer papers of Bohr and Rosenfeld \cite{BRR1}, \cite{BRR2} are still very  interesting.} Typically it is difficult  to handle the concrete measurement in the rigorous QFT-framework; often 
one has to refer to QM  which is a rough approximation of QFT having many bad  
features. As the consequence, the most intriguing problem of modern quantum foundations, the problem of interrelation 
of local realism and QT, was treated (with a very few exceptions) only in the QM-formalism. In short, QFT cannot function successfully without QM and vice versa. 
We remark that QM needs QFT to secure that QT does not contradict special relativity, that nonrelativistic character of QM is just the consequence of  the use of a convenient 
approximation, that, at least in principle,  one can always refer to the existence of the explicit relativistic theory, QFT, see appendix for further discussion.  

Now we come to the main theoretical problem of  QFT. (The absence of measurement theory is the problem of applicability and in principle there can be a gap between 
a theoretical  and observational  models.\footnote{Such ideology was advertised by creators of so called {\it  Bild conception} of the physical methodology, Hertz, Boltzmann, Schr\"odinger
(see, e.g., \cite{SCH_A1}, \cite{SCH_A2}).
Nowadays it is often present in joint handling of two levels of description of nature, ontic and epistemic, see Atmanspacher et al. \cite{H1}-\cite{H3} 
and Khrennikov  \cite{KH}, see also Jaeger \cite{Jaeger1}-\cite{Jaeger4}, \cite{Jaeger3} for discussions on quantum treatment of realism.}) This is {\it the problem of QFT-divergences.} This problem 
is well known. However, nowadays it is hidden ``under the table'': development of advanced mathematical techniques of renormalization made the impression that QFT is a consistent
theory, see, e.g., citations in appendix.  Personally I also practically forgot about this problem and in debates I always referred to QFT as the harmonically designed relative of QM. I argued that the corresponding 
foundational problems have to be lifted from QM to QFT for the proper analysis. 

Recently going back home from the exciting conference DICE 2017 (Castigliancello, Italy)  organized by T. Elze, I met a young Israeli post-doc and we started the conversation about QT 
and its merge with GR. His head was free  of  the mess of technicalities feeling my head. His questions might be naive. However, the fresh and curious insight in some problems 
can be of the great value.  His questions revoked my former thoughts about the problems of QT, both in the QM and QFT fashions; in particular, about the problem of QFT-infinities.
Since his PhD was in gravity, his questions had the strong gravitational flavor. This discussion is the basis of the presentation in section 2. It has  the form of a dialog,
questions of the naive, but fresh minded young scientist to well educated (especially in mathematical problems of QT) professor. Formally saying, we analyzed the basic problems of QM-QFT 
in connection with the problem of merging with GR. 

After this discussion my expectations for creation of new QT which will be free of all aforementioned problems and 
which will be harmonically merged with (new) GR increased to the level of strong optimism. Young generation whose fresh mind  would not accept mathematical and 
physical inconsistency of the existing theories will finally revolutionize the modern physics, cf. with citation of P. Dirac at the very   end of this paper.     
        
In principle,  after section 2 the reader can jump directly to concluding section \ref{W} (What is to be done?).  Other sections are more technical and longly.  Section \ref{MT} is about measurement theory in QM versus QFT; may be it 
can be interested for experts in quantum foundations and quantum measurement theory. Section \ref{S1} (Is Devil in HIlbert state space?) 
 enlightens the possibility to resolve at least some problems of QT by leaving the Hilbert (von Neumannian \cite{VN}) state space and proceed with a pair 
of dual linear spaces (in the spirit of P. Dirac \cite{D1}). Section \ref{S2} (Is Devil in real numbres?) 
  revokes the very long discussion about the role of numbers in physics (rational, $p$-adic, and nonstandard numbers). Section \ref{EDS} (Einstein's dream), see also \cite{ED}, reminds about Einstein's 
dreaming for the classical field unification of all physical theories \cite{EI}. This Einsteinian way of thinking is illustrated by one special model, so called {\it prequantum classical statistical field theory}
(PCSFT) \cite{Beyond}. 

\section{Problems of QT and its unification with general relativity}
\label{ST}

As was already mentioned, recently going back home from the conference DICE 2017  I was staying at the platform with  one young postdoc who asked me 
a variety of questions about QT (call this young scientist Y).  We discussed education in physics in Israel (Y is from this country). And I was positively surprised that his teacher
(who was definitely from Russia) used the books of Landau and Lifshitz. This was the good point to continue conversation.  
Since I told him that I was interested in the mathematical foundations of QFT, he suddenly asked me {\it whether 
it is really true that Feynman diagrams are divergent.} I confirmed that QFT correlations should be regularized and started to give him examples of regularization procedures. For me,  this 
was the mathematical routine. Therefore I was surprised by seeing dissatisfaction in his eyes. This was the fresh reaction of the brain which has not yet been structured 
by lecturing about the mathematical foundations of QFT. First of all Y was not happy at all that there is a variety of regularization procedures. He wanted just one and that it would be 
justified by laws of physics and not mathematics. He neither  shared my enthusiasm on the possibility to get the right  physical answers from complicated mathematical regularization procedures.
The handmade nature of QFT regularizations was really disturbing for him. 

Then it happened that Y has  already spent some time by thinking about {\it infinities} in QFT. He just was not sure that the state of art is so bad, that already so nicely sounded expression 
``Feynman diagram'' is just the symbol covering the mess of regularization procedures of QFT.  We continued our discussion at the stattion in Livorno, we still had 15 min.
For him, there are no infinities in the physical world and  infinities are just the fruits of the human mind. I agreed with him completely. Then he made the point: {\footnotesize\em ``I think that 
the appearance of infinity is just a sign that some theory is applied outside its real domain of application.''} And this  statement generated a chain of my recollections.
 
 ``Late Dirac'' was neither satisfied by modern QFT and also because the presence of infinities.\footnote{{\small In the early 1980s, P. Dirac told E. 
Witten that the most important challenge in physics was "to get rid of infinity" \cite{QFTIN}.}} This dissatisfaction was especially strongly expressed in Dirac's book \cite{D2} 
in which Dirac asked for creation of new QFT which would be free of divergences.  Y  was curious whether Dirac gave some insights towards creation of 
new divergences free QFT. My recollection was that Dirac was sure that the main problem of modern QFT is in very formal treatment of the vacuum state 
$\vert 0 \rangle.$ The real physical vacuum has a complex structure
and this structure cannot be reduced to operation with the symbol $\vert 0 \rangle.$ 

The next question was again surprising for me: {\footnotesize\em ``How could it happen that   
scientists working in QFT (and more generally QT)  did not put all their efforts to create a new divergence free QFT?''}  It seems that the answer was already in his own remark 
that infinities are well accommodated   in the human brain. People did not treat the appearance of infinities as a sign of inapplicability 
of a theory. I pointed to the Big Bang theory with its initial singularity which is also well accommodated in our brains, but, of course,
has nothing to do with physics. The Bing Bang initial singularity should be a powerful stimulus to create a new theory, e.g., in the 
spirit of the recent studies of  V. G. Gurzadyan and R.Penrose \cite{Penrose}. However, this did not happen and we peacefully live starting with  
the singular Big Bang.

Then  Y told: {\footnotesize\em ``Regarding Dirac's 
attempt to reorganize QFT by clarifying the role of vacuum, was it not just a step towards quantum gravity?''} Yes, it can be seen in this way.

And it became clear for  us both that we should not try to unify QT suffering of divergences (at the QFT level) with GR, that first we have to resolve the biggest, but 
commonly ignored (nowadays) problem of QT, namely, creation of QFT without infinities and their handmade regularizations. My next recollection was that 
infinities are not present in nonlocal versions of QFT, see e.g. \cite{QFTNL}. Thus there is some coupling between suffering from infinities and enjoying 
locality. I told that by introducing the elementary length $L$ and making impossible to go beyond it  we  can proceed without infinities. 
My friend was completely fine to live in the space which is not infinitely divisible and he preferred  such world to the QFT-world with infinities.

Then Y made the interesting point: {\footnotesize\em``See, we all know from Bell's theorem and its experimental verification that QM is nonlocal! 
Why do people have such a problem with invention of nonlocality to QFT? It is clearly nonlocal by Bell's theorem!?''} And here 
I started to present my favorite story about nonlocality of QM versus locality of  QFT:

QM is an approximation of QFT.  Do you agree? ({\footnotesize\em ``yes!''})  The real QT is QFT and not QM at all. ({\footnotesize\em ``yes!''})  
In particular, QM is nonrelativistic theory. ({\footnotesize\em ``yes!''})  There is a theory which is
called ``relativistic QM'', but it suffers of such problems that it is better to be nonrelativistic (section \ref{MT}). Therefore if one were defining 
``quantum nonlocality'' simply as the absence of relativistic invaraince, then the statement about ``nonlocality of QM'' would be trivial.
It is the good place to remark that locality of QFT  is defined precisely in the sense of relativistic invariance. 
What is about nonlocality of QM? This is a very special notion  which surprisingly does not refer to space-time at all. In particular, there 
is no space-time in the formulation of Bell's theorem. ({\footnotesize\em ``yes!''})  Nonlocality of QM is based on action at a distance which existence is seen as a consequence 
of Bell's theorem and its experimental verification.\footnote{It should be pointed out that this is a different sort of ``action'' from that which concerned, for example, 
Newton: as it cannot be used for signaling.}   Thus the notion of locality in QM and QFT have very different meanings. ({\footnotesize\em ``uf!''})
Nowadays we are excited by having the QM-nonlocality, but at the same time we are afraid of nonlocal versions of QFT. 

Y has a very good sense of logic and he was curious why nobody tried to proceed another way around and formulate the principle of action at a 
distance in the QFT framework?  Up to my knowledge such attempts have been done, but my impression 
was that one cannot proceed consistently. QFT with action at a distance seems to be as bad as QM with relativity. In particular, I recalled that 
even the notion of entanglement in QFT is not free of problems. {\footnotesize\em ``Why?''} 

One of the theoretical  problems of the QFT-treatment of  the Bell-type argument is that {\it entanglement depends on the frame.}

Of course, this is also not such a surprise,  since entanglement (per se) is a mathematical 
definition, i.e., a physical system can be separable or entangled with respect to a chosen factorization of  the  total  Hilbert space which  
describes  the  quantum  states. Indeed,  the  choice  of factorization  allows  for  pure  states to switch unitary between separability and 
entanglement.  However,  usually the experimental setup  fixes the factorization  and  applying  local  unitary transformations does  not  change  the 
entanglement properties. But, as was already stressed, measurement is properly described only in the quantum mechanical and 
QFT framework.  My impression is that  the main problem behind the QFT-treatment of 
action at a distance  is the absence of measurement theory corresponding to QFT!  Y:  {\footnotesize\em ``Well, I suspected this!''}

I finished my story by pointing that even among top experts there is not consistency in views on measurement theory in the QFT framework. 
I had a few conversations with top experimentalists in quantum optics who are also well known in quantum foundations and they agreed that 
QFT has no measurement theory and this is one of the problems of modern QT.  At the same time other top experimentalists did not see 
any specialty of QFT treatment of measurements comparing with QM (see section \ref{MT} for further discussion).

\section{Measurement theory: QM versus QFT}
\label{MT}

QM-based measurement theory started from the Born rule, then its preliminary formalization was performed by von Neumann 
and it was based on the projection postulate, finally it was formulated as theory of POVM-representable 
measurements (more generally theory of quantum instruments) in connection with theory of open quantum systems, see 
\cite{Davies}-\cite{Jaeger3} for detailed presentation of the present state of art. Theory of quantum instruments is 
mathematically rigorous and clear theory, straightforwardly  applicable to the basic quantum experiments. 

It is important to remark that  the only mathematically and conceptually sound QM is nonrelativistic QM. One may  disagree and point to relativistic QM. However, 
the latter suffers of a number of difficult problems, e.g., the problem of relativistic localization (position).  By  Hegerfeldt's theorem:  
{\it Einstein causality implies that there can be no spectral measure solution to the localization problem in relativistic QM.}

Thus, for a moment, we have the well developed theory of nonrelativistic quantum measurements. Relativistic treatment is presented by QFT-based 
measurement theory, e.g., Schweber \cite{Schweber}, Schwinger \cite{Schwinger}. My personal opinion (confirmed by conversations with a few top experts in quantum measurement theory) is that the QFT-measurement 
theory is far from the level of mathematical rigorousness and clearness approached by the QM-measurement theory. But this is just the private opinion, cf. with coming 
analysis of the viewpoints of experimentalists. 

I shall present a ``typical viewpoint'' representing very general attitude of experimentalists 
towards  quantum measurement theory, a kind of  general synthetic statement extracted from conversations during the V\"axj\"o series (2000-2016) of conferences on 
quantum foundations: 

{\footnotesize\em One may say that an experimentalist does not need  quantum measurement theory: neither in the original von Neumann form (Hermitian operators and the 
projection postulate) nor in the modern form (POVMs and quantum operations). An 
experimentalist needs  only probabilities for detection of events.  These probabilities are provided by QFT. 

In contrast to QFT,  QM is only applicable for the Minkowski 
gauge. (It works only for the flat space.)
Moreover,  the measurement theory in the QFT framework is, in fact, more simple and fundamental than the QM measurement theory 
(von Neumann's theory  or modern theory of quantum instruments). Here, in QFT, one can proceed following the ideas of
Einstein (see, e.g., Einstein and Infeld \cite{EI}): 
\begin{enumerate}
\item Only fields do exist.
\item Any interaction of fields  comes with an exchange of their excitations,  ``particles''.
\item Any measurement can be treated as the interaction on the observed field with the field of the measurement apparatus (may be even including electromagnetic field in 
experimentalist's brain). 
\item Any measurement then merely consists of counting the quantum particles arising from the interaction involving the two fields: measured input field and measurement apparatus field.
\end{enumerate}
 QFT express this counting procedure  by the measurement and calculation of the current $J,$ detected by
 the amount of `clicks' of the counting device.
This measurement of $J$ by counting clicks is exactly what in being measured every day at the LHC machine in Geneva.

Do you find in this simple scenario any need for POVM, or `entanglements'  a la Zurek, or quantum instruments a la Davies and Ozawa, or  other theoretical nasty procedures?}

It is clear that this viewpoint has some basis behind it. It has  to be analyzed in more detail than it is possible in this paper 
(which is not devoted to quantum measurement theory). For me, this position of experimenters is very close to Bohr's position: 
QT is about predictions of outputs of classical measurement devices. The first part of the ``experimentalist's statement'' about measurement theory would be supported by 
the majority of experimentalists.  The second part related to the field picture of measurement process represents the more specific viewpoint on the quantum measurement 
theory which is characteristic for some part of the experimental community.   From this viewpoint both ``systems'' and   measurement devices are represented 
by fields. The above reference to Einstein  makes the impression that  these fields are treated classically.
However, this may be not the case. In general my impression is that often experimentalists treat the (operator-valued) fields of QFT in the same manner as classical fields.

We remark that  {\footnotesize\em ``Einstein never accepted the QFT theory, because it shared all the `problems of QM',  or what he
so perceived, most especially its fundamentally nonrealist and acausal, and thus statistical nature.
In fact it is not clear to what degree he grasped or, given this character of the theory, even
wanted to grasp QFT, or QED. At one point, he developed an interest in Dirac's equation, as a
spinor equation, and he used it, in his collaborations with W. Mayer, as part of his program for
the unified field theory, but, again, conceived as a classical-like field theory, modeled on general
relativity, and in opposition to quantum mechanics and, by then, quantum field theory.
Accordingly, he only considered a classical-like spinor form of Dirac's equation, thus depriving
it of (Einstein might have thought `freeing' it from) its quantum features, most fundamentally,
discreteness (h did not figure in Einstein's form of Dirac's equation), and probability. Einstein
hoped,  but failed to derive discreteness from the underlying field-continuity. Einstein was
primarily interested in the mathematics of spinors, which he generalized in what he called
semi-vectors''} (private communication with A. Plotnitsky).

 One may say that experimentalists still 
keep to the orthodox Copenhagen interpretation, because they consider the present measurement theory as unsatisfactory. 

However, I disagree that transition from QM to QFT may simplify measurement's picture. I would like to point to the following complication due to the QFT-treatment, see also 
Plotnitsky \cite{PL16},  Chapter 6.   Since QM is a nonrelativistic model, the space-time can be, in principle, excluded from consideration. The ontological interpretation of the wave function $\psi=\psi(t,x)$
is,  although possible,  not so much supported by the recent development of quantum foundations. We can (by following Schr\"odinger) to interpret a quantum 
state as an expectations catalog.  The Schr\"odniger dynamics, $\psi_t=U_t \psi_0,$ need not be coupled to processes in physical space-time.

We remark that 
von Neumann measurement theory does not describe the process of interaction of a quantum system and a measurement device  in physical space-time. However, by the above argument 
this is fine (in the nonrelativistic  QM-framework). In QFT, physical space-time reappeared in whole its power, very similarly to classical field theory. Quantum fields are functions 
on the space-time, the ``only difference'' from classical field theory is that quantum fields are operator valued. 
The proper QFT-measurement theory has to take the space-time structure into account and to present the space-time based picture of interactions of quantum systems (fields) with 
measurement devices (represented by quantum or even classical fields).\footnote{The problem cannot be disregarded by formal transition to the particle representation 
in the Fock space. Creation and annihilation operators are still functions defined on physical space-time.} To my knowledge, there is no such a theory.  
    
It seems that by speaking about ``simply counting procedures'' we shadow thremedous difficulties of the theoretical description of these procedures
 As Pais said years ago \cite{Pais}, p. 325, but this still true:

{\footnotesize\em ``Is there a theoretical framework for describing how particles are made and how they vanish?
There is: quantum field theory. It is a language, a technique, for calculating the probabilities of
creation, annihilation, scattering of all sorts of particles: photons, electrons, positrons, protons,
mesons, others, by methods which to date invariably have the characters of successful
approximations. No rigorous expression for the probability of any of the above-mentioned
processes has ever been obtained. The same is true for the corrections, demanded by quantum
field theory, for the positions of energy levels of bound-state systems [e.g., atoms]. There is still
a [Schrödinger] equation for the hydrogen atom, but it is no longer exactly soluble in quantum
field theory.''}
 (Pais speaks of particles, but a field interpretation does not change anything in this respect.)

\section{ Is Devil in Hilbert state space?}
\label{S1}

If one reads the works P. Dirac carefully, starting with his famous book \cite{D2} and paying especial attention to his book devoted to QFT \cite{DQFT}, it becomes clear 
that P. Dirac suspected that the fundamental problems of QT, especially divergences in QFT, are consequences of the use of the Hilbert space as the mathematical model of 
the state space. In fact, he started his way to construction of the mathematical formalism of  QM not with the Hilbert state space (in contrast to von Neumann \cite{VN}), but 
with a pair of two dual linear spaces. Denote them $E_+$ and $E_-.$ The elements of one of them are called ket-vectors and of another bra-vectors; the former are denoted as 
$\vert \psi\rangle$ and the latter as $\langle f \vert.$ The duality form for these two spaces is denoted as $\langle f \vert \psi \rangle.$ 
Only later in the text of his book \cite{D2} Dirac identified these spaces $E_+ =E_-= H$ and $H$ can be mathematically treated as the Hilbert space. Of course, such identification 
simplified the mathematical formalism. The main physical argument in favor of such identification is that only in Hilbert space we can proceed with the Born rule providing 
the probabilistic interpretation of QM.  We shall discuss this statement later.

Now we point out that, in spite of  Hilbertization of his formalism, P. Dirac continued to operated 
with two types of vectors. The latter seems to be totally meaningless, cf. with the book of von Neumann \cite{VN}. However, careful analysis shows that (as from the very beginning)
P. Dirac used the ket and bra vectors as the elements of dual vector spaces. Thus elements of $E_-$ can be treated as linear functionals on the vector space $E_+.$ (These functionals 
are, in fact, the basic blocks to determine {\it observables.}) The space of ket-vectors is really the state space. However, the space of bra-vectors $E_-$ is, in fact, the space of linear 
functionals on the state space $E_+.$  

In this scheme an observable $A$ is determined by a sequence of functionals, bra-vectors, $A=(f_1,..., f_n,...).$ Each functional represents measurement 
of some  value $a_i$ of the observable $A.$ For mathematical simplicity, we consider observables with discrete ranges of values; we also suppose different functionals represent 
different values  (given by real numbers). 

Then the Born rule has the following form. For systems prepared in the state $\psi \in E_+$ and
 the observable $A=(f_1,..., f_n,...), f_j \in E_-,$ the probability to 
observe the outcome $a_i$ is given by the rule: 
\begin{equation}
\label{BR}
p(A=a_i\vert \psi)= \frac{ \vert \langle f_i \vert \psi \rangle\vert^2}{\sum_j     \vert\langle f_j\vert \psi \rangle\vert^2}.
\end{equation}
If  $\sum_j     \langle f_j\vert \psi \rangle^2 = \infty,$ then we say that the observable $A$ cannot be measured for this state.
The latter does not mean that it is something wrong with the state $\psi,$ that it is unphysical. It can be that a variety of other 
observables are measurable for $\psi.$ For some $B=(g_1,...,g_n,...),   \sum_j     \langle g_j\vert \psi \rangle^2 < \infty.$ 
This is a natural physical situation; two procedures, preparation and measurement,  should be consistent.

In the modern functional analysis, the space of ket-vectors (the state space) $E_+$ can be treated as a topological linear space, a linear space endowed with some topology such that the operations of addition 
of vectors and multiplication of scalars are continuous. The space of bra-vectors (elementary components of observables) $E_-$ is represented as the dual space, the space of continuous 
linear functionals on $E_+.$ For example, $E_+$ can be chosen as the space of Schwartz test functions $E_+={\cal S}(\bf  R^3)$ and its dual space is the space of  
Schwartz distributions $E_-={\cal S}^\prime(\bf  R^3).$ 

To make closer analogy with the Hilbert space case, consider a topological linear space $E_+$ having a (topological) basis, i.e., a system of vectors $(e_1,..., e_n,...)$ such 
that any vector $\psi$ can be expanded uniquely with respect to this system
\begin{equation}
\label{BR1}
\psi= \sum_j x_j e_j, x_j \in \bf  C,
\end{equation}
and the series converges in the topology of $E_+.$ Consider the system of linear-functionals on $E_+: (f_1,..., f_n,...), \langle f_j\vert \psi\rangle = x_j,$ i.e., 
$\langle f_j\vert e_i\rangle = \delta_{ji}.$ Then this system of functionals is a basis in the dual space $E_-.$  The latter is endowed with so called weak topology.\footnote{To make everything 
consistent, we have to consider the special class of topological linear spaces, so called locally convex spaces. In such spaces each point has the basis of convex neighborhoods.}
The systems of vectors $(e_1,..., e_n,...)$ and functionals $(f_1,..., f_n,...)$ are biorthogonal bases. The expansion  (\ref{BR1}) can be written as 
\begin{equation}
\label{BR2}
\psi= \sum_j \langle f_j\vert \psi\rangle e_j.
\end{equation}

Now we remark that we could proceed another way around: to start with a topological basis in the dual space $E_-, (f_1,..., f_n,...),$ and construct its biorthogonal basis 
in $E_+.$ The weak topology on $E_-$ has the property  that the dual space for $E_-$ coincides with $E_+.$ 

Now consider the observable $A$ given by some basis $A=(f_1,..., f_n,...).$ Let $(e_1,...,e_n,...)$ denote the corresponding basis in the state space. 
Then this observable can be represented by the linear operator $\hat A: E_+ \to E_+,$ acting as $ f \to \sum_j  a_j \langle f_j\vert \psi\rangle e_j.$ In the Dirac notation 
it can be written as 
\begin{equation}
\label{BR3}
\hat A= \sum_j  a_j \vert e_i\rangle  \langle f_j\vert.
\end{equation}
In Hilbert space we can identify the vector and functional counterparts of the system of the biorthogonal bases, set $f_j=e_j$ 
and obtain the standard spectral decomposition of a Hermitian operator.
\begin{equation}
\label{BR3}
\hat A= \sum_j  a_j \vert e_i\rangle  \langle e_j\vert.
\end{equation}

Consider now a rigged Hilbert space based on the dual pair $E_+, E_-:$ 
\begin{equation}
\label{BR4}
E_+ \subset H \subset  E_-,
\end{equation}
where $H$ is a Hilbert space and the injections of $E_+$ into  $H$ and  $H$ into  $E_-$ are continuous with respect to the topologies on these spaces.\footnote{To have a mathematically rich 
model, the topological space $E_+$ should belong to the class of so called nuclear spaces; all basic spaces of functional analysis are of such a type.}
One of common examples is based on the Schwartz spaces and the $L_2$-space:
\begin{equation}
\label{BR5}
{\cal S}(\bf  R^3) \subset L_2(\bf  R^3) \subset {\cal S}^\prime(\bf  R^3).
\end{equation}

By considering Hermitian operators in the Hilbert space $H$ we reduce essentially the class of possible observables. 
The main point of Dirac was that by considering quantum dynamical equations
in $H$  (for observables, i.e., in the Heisenberg picture)  we restrict essentially the class of possible dynamics. Infinities are induced by our attempt to represent the $E_-$-dynamics 
as the $H$-dynamics.  

In the Hilbert space model the Heisenberg and Schr\"odinger pictures are equivalent. In the general dual space model, the situation is more complicated. 
We can repeat the above scheme by selecting $E_-$ as representing the state space and $E_+$ as representing  observables. In the case of 
$E_+={\cal S}(\bf  R^3)$ and $E_-={\cal S}^\prime(\bf  R^3)$ this state-observable inversion leads to huge space of states and small space of observables.
Now a state $\psi$ is represented by an arbitrary vector $\psi \in E_-$ and an observable is represented as $A=(f_1,...,f_n,...), f_j \in E_+.$ The Born rule 
has the form: 
\begin{equation}
\label{BRST}
p(A=a_i\vert \psi)= \frac{ \vert \langle \psi \vert f_i \rangle\vert^2}{\sum_j     \vert\langle \psi\vert f_i \rangle\vert^2}.
\end{equation}
Consider again a rigged Hilbert space, see   (\ref{BR4}). Then, from Dirac's viewpoint, the main problem of QT is in attempting to restrict the state space of a quantum system 
to the fixed Hilbert space $H.$ In fact, the real state dynamics takes place not in $H,$ but in $E_-.$ In the example (\ref{BR5}), states are accommodated in the space 
of distributions and not in the $L_2$-space. 

The selection of a Hilbert space $H$ corresponds to the selection of a fixed vacuum. Dirac pointed out that the real state dynamics'
need not be concentrated in the space corresponding to this concrete vacuum (which choice corresponds to our laboratory conditions); for infinitely small instance of time
the state can be kicked out this fixed Hilbert space $H=H_0.$

If one wants to treat a quantum system as living in a Hilbert state space, then time-dependent rigged Hilbert spaces 
have to be explored:     
\begin{equation}
\label{BR4}
E_+ \subset H_t \subset  E_-,
\end{equation}

Now we remark that the example (\ref{BR5}) is only of the illustrative value, since it represents Dirac's ideology in the QM-framework, i.e., for systems 
with the finite number of degrees of freedom. To explore this scheme for QFT, one has to consider spaces of test functions and distributions on 
infinite-dimensional spaces. Non-Hilbertian approach to quantization of systems with infinite number of degrees of freedom was established by O.G. Smolyanov 
et al. \cite{SMPDO1}- \cite{SMPDO2}. 

In particular, an attempt to realize Dirac's idea about state-dynamics which is instantaneously kicked of the Hilbert state space (the Fock space) at the mathematical 
level was presented in \cite{KHRL}.  But there is still the huge gap between mathematical modeling and real physics.      

In this paper we consider constraints making difficult unification of QT and GR. The use of the Hilbert state space is one of such constraints, see, e.g.,  `t Hooft \cite{SBAR}:
{\footnotesize\em ``The standard Hilbert space procedure for quantum mechanics does not go well with gravitation, curved space time and cosmology.''} 
At the same time theory of distributions is well adapted to manifolds.  This section is aimed to remind about this important constraint. At the same time we are well aware that Dirac's own attempts to resolve quantum foundational problems 
by proceeding in the non-Hilbertian framework were not successful.

\section{Is Devil in real numbers?}
\label{S2}

We remark that the real numbers appeared in physics as a mathematical supplement of the use of Newtonian differential and integral calculus.\footnote{If physicists 
would follow Leibniz, then the modern physics were based on some version of nonstandard analysis \cite{NA1}.} The use of real numbers in physics was the culmination 
of long struggle between supporters of the ``continuous and discrete viewpoints'' on modeling of nature. 
The continuous model is rooted to the works of Aristotle. In the modern language his  ideas can be formulated as follows:
continuous geometry for the physical world and discrete geometry for the mental world. Democritus claimed that the 
world is build of atoms, {\it indistinguishable elementary blocks} of nature.\footnote{Plato hated Democritus' idea and asked to burn Democritus' writings.
We remark that, although Newton's calculi were based on infinitely divisible continuum, his model of light was of corpuscular nature.}  

Newton's idea about   infinitely divisible continuum was later elaborated mathematically and in 19th century the rigorous mathematical model of 
the field of real numbers was created. This model worked perfectly in classical mechanics and it was successfully extended to cover field theory. We remark that even theory 
of electricity was based on a kind of continuous electric fluid. Discoveries of electron and atom recovered  Democritus discrete picture of nature. Then M. Planck 
showed that the problem of black body radiation can be solved if the energy exchange between matter and radiation is modeled by using discrete 
portions of energy, quanta.  (We remark that already Boltzmann used in mathematical calculation portioning of energy by discrete ``quanta'' $\epsilon.$ Thus M. Planck 
proceeded in this direction by assigning to $\epsilon$ the concrete value $\epsilon_\nu= h \nu.)$  Einstein made the next step and claimed that energy of the radiation 
is quantized not only in the process of the energy exchange between matter and field, but the electromagnetic field by itself is quantized, in vacuum. 
We also point to Bohr's model of atom in which electrons could not move freely in the real space, but only choose the special discrete set of orbits. 
This was the right  time to look for a new mathematical model representing these discrete features of micro-world. However, 
the real numbers were peacefully incorporated in the mathematical formalism of QM. From this viewpoint, QM is an attempt 
to model discrete nature with the aid of  infinitely divisible  continuum. 

It might be that some strange features of QM and QFT, including divergences in the latter, are just the mathematical artifacts of using the real numbers model.
This viewpoint is not new and its different versions were presented by many scientists, see the monograph  \cite{KHRNA} for a discussion and references about 
this topic.  One class of proposals is based on the use of finite 
number fields, instead of the field of real numbers. Another class  is based on the idea that only rational numbers 
can be treated as ``physical numbers''.  Starting with the field of rational numbers $\bf  Q,$ one can try to construct new number fields\footnote{Algebraic
structures for which all basic arithmetic operations, addition, subtraction, multiplication, and division, are well defined.}. Surprisingly mathematics 
gives us only two possibilities to proceed in this way: either to the field of real  numbers $\bf R$ or to 
the fields of $p$-adic numbers $\bf  Q_p,$ where $p>1$ is a prime number determining the field, see, e.g.,  \cite{KHRNA}.
The latter combines the discrete and continuous features. $P$-adic analysis can be considered as the discrete analog of Newton's analysis for continuous 
entities. Mathematically $\bf Q_p$ is constructed in the same way as $\bf R,$ as completion of $\bf Q,$  but with respect to the
special $p$-adic metric. This metric is so called ultrametric; not only the standard triangle inequality, but even strong triangle inequality holds.
By the latter in each triangle the length of third side is always majorated by the maximum of the lengths of two other sides. This feature of metric on    
$\bf Q_p$ induces unusual features of $p$-adic geometry.  

Starting with the pioneer paper of I. Volovich \cite{VOL1} in 1987 (see also \cite{VOL2}, \cite{VOL3}),   
$p$-adic theoretical physics has been rapidly developing and covering all basic areas 
of physics, but with the emphasize on string theory, see, e.g., the monographs \cite{VOL4}, \cite{KHR94}, \cite{KHRNA} and the recent review \cite{R}. String theory models nature 
at the fantastically small  space and time intervals, at the Planck scale. It was natural to question the Newtonian model of space-time based on 
the field of real numbers $\bf R.$ In particular, one questioned infinite divisibility  of space-time intervals, the determining feature of the 
real continuum.  Therefore the $p$-adic space-time was welcome as one of possible candidate for the mathematical model of the Planck space-time.
The international team of researchers, Volovich, Vladimirov, Aref'eva, Witten, Dragovich, Frampton, did a lot in this direction, see \cite{R}. However, since 
string theory by itself has no direct relation to real physics, the $p$-adic string theory cannot be used to justify the use of the fields of $p$-adic numbers
$\bf Q_p$ in physics, in particular, in QFT. 

We remark that the $p$-adic theoretical physics explores two mathematical models. In both models the space is $p$-adic, i.e., instead of 
$\bf R^n,$ one uses   $\bf Q_p^n.$ 

In most developed model \cite{VOL4},  one uses the functions  $\phi: \bf Q_p^n \to \bf C,$ where 
$\bf C$ is the ordinary field of complex numbers. For example, in corresponding $p$-adic QM wave functions are 
complex valued, as in the standard QM. The state space is the space of square integrable functions (with respect to the Haar measure on $\bf Q_p^n), H=L_2(\bf Q_p^n),$
and this is the usual complex Hilbert space. Spectra of operators representing observables are subsets of $\bf R.$  The standard 
probability interpretation, the Born rule, can be used. The main difference from standard QM is that in $p$-adic QM it is impossible 
to construct an analog of operators of position and momentum, a kind of Heisenberg algebra.\footnote{The standard Schr\"odinger representation  
of $\hat x_j$ and $\hat p_j$ based on the multiplication and differentiations cannot be extended to $p$-adic QM by the purely mathematical reason:
it is impossible to multiply $p$-adic and real (complex) numbers. One cannot multiply  a function  $\phi: \bf Q_p^n \to \bf C$ by 
one of its variables $x_j \in \bf Q_p.$ This also implies that it is impossible to define derivatives of such functions \cite{KHR94}.}
Mathematically the $p$-adic model with $\bf C$-valued functions can be treated as a part of harmonic analysis on locally compact groups 
(the fields of $p$-adic numbers are locally compact additive groups).\footnote{However, the presence of the field structure, i.e., combination 
of the operations of addition and multiplication, and special features of the distance on $\bf Q_p$ which is ultrametric make $p$-adic 
analysis essentially richer and it has some very special features which are not common for locally compact groups in general.}  

In another model of $p$-adic theoretical physics  (see \cite{KHR94}, \cite{KHRNA}, \cite{KHRUSP}-\cite{PADIC}),  
both variables and values are $p$-adic.\footnote{ The latter can belong to quadratic and more general algebraic 
extensions of $\bf Q_p,$ analogs of the field of complex numbers $\bf C$ which is the quadratic extension of $\bf R,\; \bf C= \bf R(\sqrt{-1}).$}
This model is farer from standard QM than the $[Q_p^n \to \bf C]$-model.   The main mathematical problem of the ``genuine $p$-adic model'' 
is that the analog of the Born rule leads to $p$-adic quantities having the meaning of probabilities. This is the good place to remark that real numbers 
infiltrated into physics much deeper, not only space-time coordinates, but even probabilities are represented by real numbers. In a series of works 
of the author (see, e.g., \cite{KHR94}, \cite{KHRNA},  \cite{KHRP3}),  there was designed probability theory with $p$-adic valued probabilities. 
They are defined in the spirit of von Mises frequency theory 
of probability, as the limits of frequencies, 
\begin{equation}
\label{BRSTj}
P(a)= \lim_{N\to \infty} \frac{n_a}{N},
\end{equation}
where $n_a$ is the number of observations of the result $a$ and $N$ is the total number of observations. Here the limit is taken with respect to the $p$-adic metric.
We remark that frequencies  always belong to the field of rational numbers $\bf Q.$ This field can be embedded both in the fields real and $p$-adic numbers, 
$\bf Q \subset \bf R$ and $\bf Q \subset \bf Q_p$ for any $p>1.$ Therefore any sequence of frequencies  $\{q_N(a)=\frac{n_a}{N} \}$can be handled both as a sequence in 
$\bf R$ and in $\bf Q_p.$ $P$-adic probabilities are defined for sequences having the limits in  $\bf Q_p.$

This theory questioned applicability of the standard law of large numbers. The existence of the limit 
 (\ref{BRSTj}) in $\bf Q_p,$ typically implies nonexistence of such a limit in $\bf R.$  \footnote{We remark that the standard ``real law of large numbers''
is, in fact, the basis of the modern scientific methodology: only observations satisfying it are considered as scientifically meaningful.  Statistical data violating 
this law is considered as output of badly performed observations.} Thus reconsideration of QT on the basis  the ``genuine $p$-adic model''  should lead 
to reconsideration of the role and interpretation of probability in QT. The first steps in this direction were done in author's works   \cite{KHR94}, \cite{KHRNA}, 
\cite{KHREPR} (the latter paper is devoted to the $p$-adic probabilistic treatment of the EPR-experiment). However, these studies
are still of purely theoretical nature. 

We also point to the $p$-adic attempts  to beat quantum divergences by considering $p$-adic quantities, instead of the real ones (see, e.g., the papers of  Cianci and Khrennikov
\cite{PDIV1}, \cite{PDIV2}).

A very different program of reconsideration of QT  through playing with the basic number field is based on the Leibniz approach to calculus. We recall that he interpreted 
infinitely small and large quantities on the same grounds as finite ones. In   Leibniz's calculus one can operate with, e.g., infinitely large numbers, add them to finite numbers, 
subtract  them, multiply, and divide. It is very attractive to apply the Leibniz approach to handle QFT infinities. One of the rigorous mathematical models 
for Leibniz's calculus is based on the field of {\it nonstandard numbers} (see, e.g., Robinson's book \cite{NA1}).
 It is typically denoted by the symbol $^\star\bf R.$ It is an extension of 
$\bf R$ - to include infinitely small and large quantities.\footnote{The $p$-adic program of reconsideration of QT (and physics in general)
is based on the idea that a kind of the discrete model has to serve as the mathematical basis of such reconsideration. The nonstandard program 
is based on the idea that one should not afraid infinities at all, accept them as physical quantities. Thus the $p$-adic and nonstandard programs question 
the standard real physics, but from the opposite sides. On the other hand, we can mention the common feature of the $p$-adic and nonstandard 
fields. They both are {\it non-Archimedean fields.} The Archimedean axiom is violated. This axiom can be treated (due to Volovich \cite{VOL1}) as 
a measurement axiom. In the Archimedean measurement theory by having two quantities $l$ (e.g., unit of measurement) and $L$ 
it is always possible to measure $L$ (at least approximately) with the aid of $l.$  In the non-Archimedean theory this natural 
feature of the  measurement process can be violated.}  The applications of nonstandard analysis to QFT were considered by a few authors \cite{NA2}-\cite{NA5}.\footnote{I was lucky
to cooperate closely with S. Albeverio who did a lot for establishing of the nonstandard approach of QFT, e.g., \cite{NA4}, \cite{NA5}. In early 1990th he was very enthusiastic and his expectations for nonstandard 
reconsideration of QFT were really great. It seems that later he lost the interest to this program. (It may be that his later interest to the 
$p$-adic approach, see, e.g., \cite{PAL}, \cite{PADIC} was stimulated by disillusion in possibilities of the nonstandard approach.) In general nowadays the nonstandard analysis is practically dead, only a very few people
still try to use it.}  

In short, development of $p$-adic physics (based on consideration of only rational numbers as physical numbers) can be characterized as very successful  theoretically, but 
its models are still far from experimental verification. The same can be said about nonstandard physical models. 
Nevertheless, it is useful to keep in mind this way  of reconsideration of QT (i.e., exclusion of real numbers from quantum physics and playing with other number fields and more general 
algebraic structures).
    
\section{Einstein's dream}
\label{EDS}

By the Copenhagen interpretation of QM this theory is complete, i.e., it is impossible  to find a finer description of physical phenomena in the micro-world than  the 
wave-function description.\footnote{ See, e.g., Bohr's reply \cite{BR0} to the EPR-paper \cite{EPR}:
``In this connection a viewpoint termed ``complementarity'' is explained 
from which quantum-mechanical description of physical phenomena would seem to fulfill, within its scope, all rational demands of completeness.''} For the fathers of this interpretation, Bohr, Heisenberg, Pauli and Fock, completeness was the straightforward consequence of the existence    of the fundamental quantum 
of action, given by the Planck constant $h.$ They did not need to search for no-go theorems. The Heisenberg uncertainty relation was the deciding ``no-go theorem''. 
The completeness of QM implies that it is impossible to emerge it from some subquantum classical-like (causal) theory (see, e.g.,  Jaeger \cite{Jaeger1}-\cite{Jaeger4}, \cite{Jaeger3} for detailed 
analysis). 

Nowadays a variety of no-go statements prevents  treatment of QM as emergent theory. Moreover, nowadays the really misleading element, quantum nonlocality, plays 
the crucial role in debates on the possibility to emerge QM from a classical-like model,  \cite{B1}, \cite{B2}, \cite{W}.  It is widely claimed that such emergence 
is possible only by assuming the presence 
of nonlocal interactions of the form of the Bohm quantum potential. Bohmian-type emergence of QM is totally foreign to GR and the general Einstein 
attitude to reproduce QM/QFT from a (local)  classical field model, see Einstein and Infeld \cite{EI}.    Such emergence cannot serve to unification of QT and GR,
since the latter is local theory.

However, by distancing from statements such as the Bell no-go theorem we can consider the following plan for unification of QT with GR \cite{KH}:
\begin{itemize}
\item Unification of all physical interactions in a single  classical field model ${\cal M}.$ 
\item Emergence of QT from  ${\cal M}.$
\end{itemize}
Of course, ${\cal M}$ has to be local and to have nothing to do with nonlocal subquantum models of the Bohmian type.

One may immediately say that such ${\cal M}$ does not exist as the result of the experimental confirmation of a violation of the Bell inequality. 
However, this conclusion depends on treatments of possible ways of coupling a subquantum (ontic) model with QM (epistemic model), see
\cite{KH} for discussion. In short, a violation of the Bell inequality closed only one special way for coupling of subquantum and quantum models.
One  can still  treat QM as emergent from a causal classical-like model ${\cal M}$ \cite{Beyond}.

For us, it is important  that  ${\cal M}$ can be a classical field model. One of such models was developed in the series of the works 
of the author and coauthors \cite{KH1}-\cite{TSD6}, so called {\it prequantum classical statistical field theory} (PCSFT), see monograph 
\cite{Beyond} for the detailed presentation. The idea behind this model and
emergence of QM is very simple: in fact, quantum density operators are just the symbolic representations of covariance operators 
of classical random fields fluctuating at the time and space scales which are essentially finer than the scales of quantum experimental physics.
We are not able to monitor individual fluctuations; only correlations can be measured. 

There is a kind of nonlocal element in PCSFT. However, this is not nonlocality of mystical action at a distance, but the presence of correlations 
in the common background field feeling the space-time (similar to the zero-point field). Its correlations contribute into the ``quantum correlations.'' We repeat once again that 
correlations in such a background field are typical for, e.g., radio-engineering.

\section{What is to be done?}
\label{W}

Now I would like to summarize the discussion with student, see section \ref{ST}, and my own reflections induced by this discussion, sections \ref{S1}-\ref{EDS}:  

\begin{enumerate}
\item Both QM and QFT were just steps towards real QT which have not yet been constructed.
\item Nonrelativistic QM versus relativistic QFT. 
\item QFT suffers of infinities. Ignorance of infinities under the shadow of regularization procedures cannot be considered as acceptable.
\item Infinities and singularities  are foreign to real physical phenomena. 
A theory containing singularities cannot be considered as physically adequate, neither QFT nor Big Bang. 
\item Since not QM, but QFT is considered as fundamental QT-model, the basic output of foundational studies in the QM framework, namely, the foundational 
output of violation of Bell's inequality - existence of action at a distance, has to be translated to the QFT-language.
\item The latter seems to be impossible, since there is no the QFT-based  notion of entanglement, it is frame dependent.
\item Unification  of QT with GR is a very risky adventure, because the real QT has not yet been created. 
Its initially created part in the form of QM suffers of the absence of relativistic invariance, it advanced part, QFT, suffers of infinities  and problems with 
relativistic notion of entanglement. 
\end{enumerate}       

Our analysis of the basic problems of QT (QM-QFT) and corresponding complications preventing unification of QT and GR has not to lead to pessimistic conclusions.
Following to one of the first Russian revolutionaries, Nikolay Chernyshevsky \cite{CH}, we can ask:  {\it ``What is to be done?''}

\begin{enumerate}
\item To create QFT which will be free of infinities (may be by following Dirac \cite{D2}).

\item  If elimination of infinities were possible only through violation of relativistic invariance of QFT, 
we have to estimate consequences of such a change for the physics in whole. 

\item We may hope that in such a new QT based on divergence free QFT action at a distance would get a new interpretation.

\item The use of real (and, hence, complex) numbers and Hilbert space are very strong mathematical constraints. Physical justification 
of their use can be questioned.  

\item Hilbert state space  is so foreign to gravity, curved spaces and cosmology (see `t Hooft \cite{SBAR}) that it seems to be natural to try 
to explore non-Hilbertian models in the spirit of P. Dirac   to proceed to unification of QT with GR. 

\item In models  based on non-Archimedean space the problem of (non-)locality is represented in the new way. There is not so much difference between
classical and quantum models, they both are nonlocal from the viewpoint of the real metric.   

\item  By rejecting the impossibility to treat QM-QFT as emergent from classical-like causal theories we open new way to merge 
QT and GR through unification at the level of classical field theory. This way can be called ``Einstein's dream way'', see \cite{EI},\cite{ED}.
Our PCSFT-model can serve as one of possible candidates for such a unification.

\end{enumerate}

We finish our paper by the famous citation of P. Dirac \cite{QFTD}, p. 85: 

\medskip

{\footnotesize\em ``It seems clear that the present quantum mechanics is not in its final form. Some further changes will be needed, just about as drastic as the changes made in passing 
from Bohr's orbit theory to quantum mechanics. Some day a new quantum mechanics, a relativistic one, will be discovered, in which we will not have these infinities 
occurring at all. It might very well be that the new quantum mechanics will have determinism in the way that Einstein wanted.''}\footnote{And, as we emphasized in this note, 
 such a divergence free and relativistic quantum theory would be able to merge peacefully with general relativity.}

  \section*{Acknowledgments}
  
 This paper was supported by the grant of Faculty of Technology of Linnaeus University ``Mathematical modeling of complex hierarchic systems'' and 
 the grant of the EU-network QUARTZ. 
  
 \section{Appendix} 

{\bf  1. QM and special relativity theory (SRT).} {\footnotesize Relativistic Quantum Field Theory is a mathematical scheme to describe the sub-atomic particles and forces. 
The basic starting point is that the axioms of Special Relativity on the one hand and those of Quantum Mechanics on the other, should be combined into one theory. ...
Since the energies available in sub-atomic interactions are comparable to, and often
larger than, the rest mass energy $mc^2$ of these particles, they often travel with velocities
close to that of light, $c,$ and so relativistic effects will also be important. Thus, in the
first half of the twentieth century, the question was asked: \em{How should one reconcile Quantum Mechanics with Einstein's theory of Special Relativity?}  
Quantum Field Theory is the answer to this question.''} (See `t Hooft \cite{HQFT1}.

 \footnotesize{``One can say that QFT results from the successful reconciliation of QM and SRT. 
... There is also a manifest contradiction between QM and SRT on the level of the dynamics. 
The Schrödinger equation, i.e. the fundamental law for the temporal evolution of the quantum mechanical state function, 
cannot possibly obey the relativistic requirement that all physical laws of nature be invariant under Lorentz transformations. The Klein-Gordon and Dirac equations, resulting from the search for relativistic analogues of the Schrödinger equation in the 1920s, do respect the requirement of Lorentz invariance. 
Nevertheless, ultimately they are not satisfactory because they do not permit a description of fields in a principled quantum-mechanical way.''} (See Kuhlmann \cite{Kuhlmann}.)

{\bf  2. QFT-foundational studies.} As was rightly pointed out by one the reviewers of this paper:   {\footnotesize\em 
``Apart from the problem of the unification or compatibility with GR, ... QFT  is far
from complete or unified. Two main parts of the  Standard Model of the
elementary particle physics, the electroweak theory, unifying quantum electrodynamics
(QED) and the theory of weak interaction, and quantum chromodynamics (QCD), which
handles the strong interactions, are fully compatible but not really unified (there is no
single established symmetry group). There are also a variety of foundational approaches
to QFT in the first place.''}  Some of these approaches are discussed  in Kuhlmann's review article \cite{Kuhlmann}. 
And the main stream of  QFT-foundational studies is directed towards the resolution of the special QFT interpretational problems, e.g., 
particle versus field interpretations of QFT. However, as was mentioned in introduction, a few authors contributed to quantum information related 
problems of the QFT-foundations, see, e.g., D' Ariano \cite{DA1},  Bisio, D' Ariano and Perinotti  \cite{DA2},  and Plotnitsky \cite{PD}-\cite{PD1}.
 
{\bf  3. QFT-divergences.} It was again rightly pointed out by the same reviewer: {\footnotesize\em  ``Certainly, the question of divergences and renormalization has been discussed and
the presence of divergences has seen as problems throughout the history of QFT. Such
important (more) recent developments as renormalization group and effective QFTs were in part responding to these problems.''}  
As a consequence, nowadays practically nobody claims that QFT should be rejected for manipulating with nonphysical (infinite) entities, cf. with 
 `t Hooft's historical account of development of QFT \cite{HQFT}, p. 3: 
{\footnotesize ``Before 1970, the particle physics community was (unequally) divided concerning the relevance
of quantized fields for the understanding of subatomic particles and their interactions.
On hindsight, one can see clearly why the experts were negative about this
approach. Foremost was the general feeling that this theory was ugly, requiring various fixes to cover up its internal mathematical inconsistencies.''} 
Then, see \cite{HQFT}, pp. 3-21, he continues to describe further development of QFT as successful (although difficult) resolution of all these inconsistencies.\footnote{This  
great project was based on the contributions  of  Kramers, Schwinger, Dyson, Tomonaga, Feynman, Landau, Gell-Mann and  Low, ..., 
De Witt, Faddeev, Popov, Feynman and Stanley Mandelstam, Slavnov, Taylor, ...., Englert, Brout, Higgs and Weinberg,...., Veltman, and, `t Hooft by himself, 
see  \cite{HQFT}, pp. 22-25, for the corresponding references.} Consequently nowadays there is no this general feeling that QFT is an ugly and inconsistent theory
(cf. with Dirac \cite{D2}, \cite{QFTD} and with the above citation of `t Hooft).

\end{document}